\documentclass[prl,aps,twocolumn,superscriptaddress]{revtex4-1}
\usepackage{amssymb,amsfonts,amsmath,amstext,graphicx,hyperref}

\hypersetup{
	colorlinks=true,
	linkcolor=blue,
	filecolor=magenta,      
	urlcolor=cyan,
}
\usepackage[dvipsnames]{xcolor}

{\catcode`\|=\active
  \gdef\Braket#1{\begingroup
\mathcode`\|32768\let|\BraVert\left<{#1}\right>\endgroup}}
\def\BraVert{\egroup\,\mid\,\bgroup}

\begin{document}

\title{Squeezing Enhances Quantum Synchronization}

\author{Sameer Sonar}
\affiliation{Department of Physics, Indian Institute of Technology-Bombay, Powai, Mumbai 400076, India}

\author{Michal Hajdu\v{s}ek}
\email{cqtmich@nus.edu.sg}
\affiliation{Centre for Quantum Technologies, National University of Singapore, 3 Science Drive 2, 117543 Singapore, Singapore}

\author{Manas Mukherjee}
\affiliation{Centre for Quantum Technologies, National University of Singapore, 3 Science Drive 2, 117543 Singapore, Singapore}
\affiliation{Department of Physics, National University Singapore, Singapore 117551}
\affiliation{MajuLab, CNRS-UNS-NUS-NTU International Joint Research Unit, UMI 3654, Singapore}

\author{Rosario Fazio}
\affiliation{ICTP, Strada Costiera 11, 34151 Trieste, Italy}
\affiliation{NEST, Scuola Normale Superiore \& Instituto Nanoscienze-CNR, I-56126 Pisa, Italy}
\affiliation{Centre for Quantum Technologies, National University of Singapore, 3 Science Drive 2, 117543 Singapore, Singapore}

\author{Vlatko Vedral}
\affiliation{Department of Physics, University of Oxford, Parks Road, Oxford, OX1 3PU, UK}
\affiliation{Centre for Quantum Technologies, National University of Singapore, 3 Science Drive 2, 117543 Singapore, Singapore}

\author{Sai Vinjanampathy}
\email{sai@phy.iitb.ac.in}
\affiliation{Department of Physics, Indian Institute of Technology-Bombay, Powai, Mumbai 400076, India}
\affiliation{Centre for Quantum Technologies, National University of Singapore, 3 Science Drive 2, 117543 Singapore, Singapore}

\author{Leong-Chuan Kwek}
\affiliation{Centre for Quantum Technologies, National University of Singapore, 3 Science Drive 2, 117543 Singapore, Singapore}
\affiliation{Institute of Advanced Studies, Nanyang Technological University, Singapore 639673}
\affiliation{National Institute of Education, Nanyang Technological University, Singapore 637616}

\date{\today}

\begin{abstract}
It is desirable to observe synchronization of quantum systems in the quantum regime, defined by low number of excitations and a highly non-classical steady state of the self-sustained oscillator.
Several existing proposals of observing synchronization in the quantum regime suffer from the fact that the noise statistics overwhelms synchronization in this regime.
Here we resolve this issue by driving a self-sustained oscillator with a squeezing Hamiltonian instead of a harmonic drive and analyze this system in the classical and quantum regime.
We demonstrate that strong entrainment is possible for small values of squeezing, and in this regime the states are non-classical.
Furthermore, we show that the quality of synchronization measured by the FWHM of the power spectrum is enhanced with squeezing. 
\end{abstract} 
\maketitle

\makeatletter

\textit{Introduction.---}
Synchronization is a ubiquitous phenomenon observed in a plethora of vastly different scenarios and has been extensively studied in both naturally occurring as well as engineered systems \cite{pikovsky2003synchronization,balanov2008synchronization,strogatz2004sync,kuramoto2012chemical}.
At its core it can be viewed as adjustment of rhythms of self-sustaining or chaotic systems due to either external drive or mutual coupling between the systems \cite{strogatz2014nonlinear,jenkins2013self}.
Recent years have seen a growing interest in synchronization phenomena in the quantum regime \cite{mari2013measures,ameri2015mutual}.
Phase-locking has been studied in driven quantum self-sustained oscillators \cite{marquardt2006dynamical,zhirov2008synchronization,walter2014quantum,lee2013quantum,amitai2017synchronization} while several interacting oscillators were shown to adjust their phase relationship in a manner analogous to classical systems\cite{lee2014entanglement,morgan2015oscillation,walter2015quantum,lee2013quantumcorr,giorgi2013spontaneous,manzano2013synchronization,weiss2016noise,bemani2017sync}.

In order to access the quantum regime in these proposed implementations, strong nonlinear damping rates are desired in order to obtain steady states with low average populations.
In the experiments carried out in nanomechanical resonators \cite{shim2007synchronized,bagheri2013photonic,matheny2014phase}, micromechanical \cite{zhang2012synchronization,zhang2015synchronization} and optomechanical \cite{heinrich2011collective} oscillators, a common drawback was that the system under investigation was highly excited.
This in turn limited our ability to witness any genuinely quantum effects. New implementations of self-sustained oscillators operating deeply in the quantum regime were proposed recently \cite{hush2015spin,nigg2017observing}.

In this new regime quantum fluctuations play a much more prominent role and in fact hinder the systems ability to synchronize to an external drive by introducing a new source of phase diffusion into the system \cite{walter2014quantum}.
At first glance this seems to disqualify systems operating near the ground state from being suitable candidates for the study of synchronization.
We show that this is not necessarily the case and that the complications associated with added noise originating from quantum fluctuations can be overcome with another quintessential quantum effect, namely squeezing.

In this manuscript, we show that squeezing can produce (a) stronger synchronization, (b) a narrower observed steady-state power spectrum, defined by $S(\omega)=\int_{-\infty}^{\infty}d\tau e^{i\omega\tau}\langle\hat{b}^{\dagger}(\tau)\hat{b}(0)\rangle_{ss}$, and (c) steady states that are genuinely non-classical. By analyzing squeezing as an effective two-photon drive, we show that the mean of the observed power spectrum is closer to the target frequency and  the FWHM of the power spectrum is smaller in comparison to the external drive considered in the literature. By replacing the external drive with a squeezing Hamiltonian, we overcome the deleterious effects of noise and open up the potential of observing quantum synchronization in the deep quantum regime. 

We begin by a brief overview of classical and quantum van der Pol oscillator driven by an external harmonic drive and its synchronization properties.
We then introduce our model and analyze the classical bifurcation in the generic case when both the external harmonic drive and squeezing are present.
We compare the classical phase-space behavior with the steady-state Wigner function in the quantum regime.
After this, we focus on the cases when either the harmonic drive or the squeezing is present but not both at the same time in order to better contrast their properties in the quantum regime and demonstrate that squeezing is more effective at entraining the van der Pol oscillator.
Finally, we discuss two implementations of our model, one using trapped ions and the other using optomechanics, highlighting potential applications of our results to emerging quantum technologies.

\textit{Van der Pol oscillator.---}
The van der Pol (vdP) oscillator \cite{van1920theory} driven by an external harmonic drive is given by
\begin{equation}
	\ddot{x} - \mu(1-x^2)\dot{x} + \omega_0^2x = F\cos(\omega_dt).
	\label{eq:vdp_classical}
\end{equation}
Here $\omega_0$ is the natural frequency of the vdP, $\omega_d$ is the frequency of the drive with strength $F$, and non-linearity $\mu$.
As expected, if the detuning $\Delta=\omega_0-\omega_d$ is too large for a given driving strength $F$, the oscillator does not synchronize to the drive. In phase space this can be seen as a limit cycle enclosing the origin suggesting that the oscillator does not develop a preferred phase. The situation changes for small enough $\Delta$, when a stable fixed point emerges in the phase plane meaning that the phase difference attains a fixed value and the oscillator becomes phase-locked to the drive and oscillates at the frequency $\omega_d$ \footnote{For large driving $F$ it is possible for a limit cycle to develop that does not enclose the origin meaning the absolute value of the phase difference is not constant yet is bounded. This is known as suppression of natural dynamics \cite{pikovsky2003synchronization,balanov2008synchronization} and is a synchronisation phenomenon different from phase-locking.}.

Quantum equivalent of the driven vdP is given by the following master equation in the frame rotating with the external drive \cite{lee2013quantum,walter2014quantum},
\begin{equation}
	\dot{\rho} = -i[ \Delta\hat{b}^{\dagger}\hat{b} + iF(\hat{b} - \hat{b}^{\dagger}),\rho ] + \gamma_1\mathcal{D}[\hat{b}^{\dagger}]\rho + \gamma_2\mathcal{D}[\hat{b}^2]\rho.
	\label{eq:vdp_quantum}
\end{equation}
Here $\mathcal{D}[\hat{O}]\rho = \hat{O}\rho\hat{O}^\dagger - \{\hat{O}^\dagger\hat{O},\rho\}/2$ represents Lindblad evolution, $\gamma_1$ and $\gamma_2$ are rates for linear pumping and nonlinear damping, respectively.
In the limit of the system being highly populated $(\gamma_1\gg\gamma_2)$, Eq.~(\ref{eq:vdp_quantum}) reproduces \footnote{ if the equations are classically truncated} the equation of motion derived from Eq.~(\ref{eq:vdp_classical}) in the weak nonlinearity limit $(\mu\ll1)$. Recently, a microscopic derivation of this evolution was presented in \cite{chia2017quantum}.

The system described by Eq.~(\ref{eq:vdp_quantum}) is different from the classical case of Eq.~(\ref{eq:vdp_classical}) owing to the uncertainty principle. This is reflected in the behaviour of phase-space quasi-probability distributions like the Wigner function, which is centered about the classical fixed points.  In \cite{walter2014quantum} the authors demonstrated that in the deep quantum regime given by $\gamma_1\ll \gamma_2$, the power spectrum widens considerably about the classical value. This means that though there is synchronization in this deep quantum regime, given by low number of photons, the quality of synchronization worsens.

\textit{Squeezed vdP.---} The squeezing Hamiltonian for degenerate parametric down conversion process is given by $H_{sq}=i\chi^{(2)}(\hat{b}^2\hat{c}^\dagger-\hat{b}^{\dagger 2}\hat{c})$ \cite{gerryintroductory}, where $\hat{c}$ is the pump mode, $\hat{b}$ is the signal mode and $\chi^{(2)}$ is the second-order nonlinear susceptibility.
We make the standard parametric approximation whereby we assume that the pump mode depletion is negligible and approximate $\hat{c}$ by $\lambda \exp{-i(\omega_pt-\theta)}$.
When $\omega_p=2\omega_d$ the total Hamiltonian in the frame rotating at frequency $\omega_d$ is
\begin{align}
	\hat{H}_{tot}=\Delta\hat{b}^\dagger\hat{b}+iF(\hat{b}-\hat{b}^\dagger)+i\eta (\hat{b}^2e^{-i\theta}-\hat{b}^{\dagger2}e^{i\theta}),
\label{sq_H}
\end{align}
where $\eta=\chi^{(2)}\lambda$ is the squeezing parameter. Including the standard terms for linear pumping and nonlinear damping, the full master equation is given by
\begin{equation}
	\dot{\rho} = -i\left[ \hat{H}_{tot},\rho \right] + \gamma_1\mathcal{D}[\hat{b}^\dagger]\rho + \gamma_2\mathcal{D}[\hat{b}^2]\rho.
	\label{eq:master_eq}
\end{equation}
Eq.~(\ref{eq:master_eq}) has two contrasting regimes. When $\eta=0$ and $F\neq0$, Eq.~(\ref{eq:master_eq}) reduces to a \textit{harmonically-driven} vdP usually considered in literature \cite{lee2013quantum,walter2014quantum,lorch2016genuine}.
When $\eta\neq0$ and $F=0$, we obtain a previously unexplored regime which we refer to as \textit{squeezing-driven} vdP.
We note that in \cite{weiss2017quantum} the authors considered a linearization of the harmonically-driven vdP and showed that the nonlinear model of Eq.~(\ref{eq:vdp_quantum}) can be approximated by an effective squeezing Hamiltonian.
This is a very different scenario to ours as we are interested in the effects of squeezing in the regime where linearization is not applicable.

\begin{figure}[t]
	\includegraphics[width=\columnwidth]{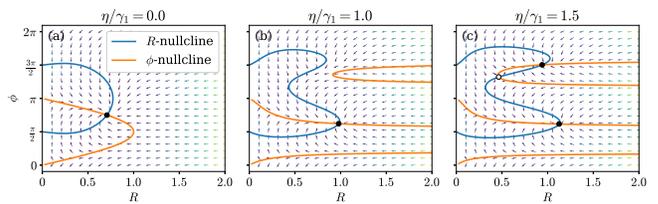}
	\caption{\label{fig:figure1} \textit{Classical phase plane diagram.}  (a) $\eta/\gamma_1=0$, (b) $\eta/\gamma_1=1$, and (c) $\eta/\gamma_1=1.5$, the blue and orange curves show $R$- and $\phi$-nullclines, respectively. For small squeezing parameter only a single fixed point exists (solid black circle) while for large enough $\eta/\gamma_1$ two new fixed points are created, one unstable (empty white circle) and one stable  as displayed in (c) and discussed in the main text. The other parameters are $F/\gamma_1=1$, $\Delta/\gamma_1=1$, $\theta=\pi/4$ and $\gamma_2/\gamma_1=3$.}
\end{figure}

To gain intuition of the fixed points of the dynamics given by Eq.~(\ref{eq:master_eq}), we begin by deriving the classical equations of motion. When the oscillator is highly excited ($\gamma_1\gg\gamma_2$) we can replace operator $\hat{b}$ with its average, $\langle\hat{b}\rangle:=Re^{i\phi}$, leading to the following coupled system of equations
\begin{align}
	\dot{R} & = \frac{\gamma_1}{2}R - \gamma_2R^3 - F\cos\phi - 2\eta R\cos(2\phi-\theta), \label{eq:R_dot} \\
	\dot{\phi} & = -\Delta + \frac{F}{R}\sin\phi + 2\eta\sin(2\phi-\theta). \label{eq:phi_dot}
\end{align}
In the simple case of driving on resonance ($\Delta=0$) and squeezing along the position quadrature ($\theta=0$), Eq.~(\ref{eq:phi_dot}) displays a pitchfork bifurcation \cite{strogatz2014nonlinear}.
This can be seen by looking at the dynamical equation for phase $\phi$ obtained from Eq.~(\ref{eq:phi_dot}) and setting the time derivative to zero, $0=\sin\phi_{ss}\left(F/R_{ss}+4\eta\cos\phi_{ss}\right)$.
For small squeezing parameter $\eta$ only a single stable fixed point exists at $\phi=\pi$.
As $\eta$ increases $\phi=\pi$ becomes unstable and two new stable fixed points, symmetric about $\phi=\pi$, emerge.
This symmetry is broken for finite detuning.
Now a single fixed point exists for larger values of $\eta$ compared to the resonant case and when bifurcation finally occurs the two stable solutions are no longer symmetric about $\phi=\pi$.
This is summarized in Fig.~\ref{fig:figure1}.
\begin{figure}[t]
	\includegraphics[width=\columnwidth]{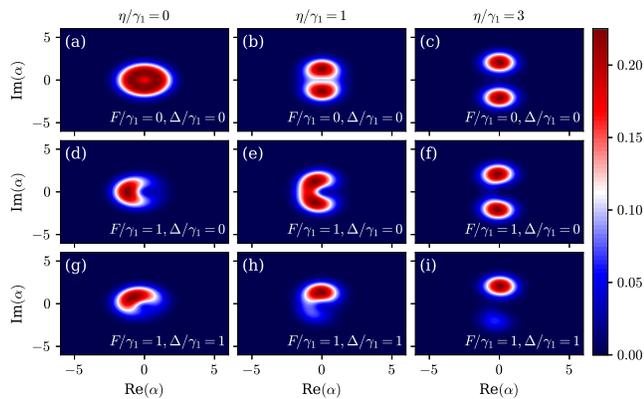}
	\caption{\label{fig:figure2} \textit{Bifurcation of the Wigner function}. For column (a)-(g) $\eta/\gamma_1=0$, for column (b)-(h) $\eta/\gamma_1=1$ and for column (c)-(i) $\eta/\gamma_1=3$. Row (a)-(c): Undriven vdP when $F/\gamma_1=0$, $\Delta/\gamma_1=0$. Squeezing acts to split the Wigner function into two localized lobes symmetric around $\text{Im}(\alpha)=0$. Row (d)-(f): $F/\gamma_1=1$, $\Delta/\gamma_1=0$. Similar to the undriven case, the oscillator displays symmetric bifurcation with increasing $\eta$. The difference is that the oscillator develops a definite preferred phase when $\eta/\gamma_1=0$. Row (g)-(i): $F/\gamma_1=1$, $\Delta/\gamma_1=1$. Detuning breaks the symmetry of the above two cases as one of the lobes of the Wigner function nearly completely vanishes. All plots are in the regime of few excitations $\gamma_2/\gamma_1=3$, and squeezing is along the position quadrature $\theta=0$.}
\end{figure}

This behavior is observed also in the regime when the average population of the oscillator is close to the ground state.
The steady-state solution $\rho_{ss}$ of Eq.~(\ref{eq:master_eq}) is obtained numerically \cite{johansson2012qutip,johansson2013qutip} for different values of $\eta$, $F$ and $\Delta$ and the steady-state Wigner functions are plotted in Fig.~\ref{fig:figure2}. The bifurcation behavior observed from the classical solutions can be identified as splitting of the Wigner function into two symmetric parts when driven on resonance. For finite detuning $\Delta$ this symmetry is broken as can be seen by the lowering of one of the Wigner function peaks. This is consistent with the observation of bifurcation of the phase distribution associated with squeezed vacuum \cite{schleich1989bifurcation}.

\emph{Synchronization without external drive.---} Squeezing may be viewed as a two photon drive suggesting that the harmonic drive in Eq.~(\ref{eq:master_eq}) is not necessary in the presence of non-zero squeezing. To investigate this, we note that Eq.~(\ref{eq:phi_dot}) has a stable steady-state solution even in the absence of the harmonic drive, meaning the oscillator becomes phase-locked and entrained to frequency $\omega_d$.
The stable solution
\begin{equation}
	\phi_{ss} = \frac{1}{2}(\pi+\theta) -\frac{1}{2}\sin^{-1}\left(\frac{\Delta}{2\eta}\right)
\end{equation}
exists provided $\eta\geq|\Delta|/2$.
The phase of the pump mode $\theta$ rotates the solution $\phi_{ss}$ in phase-space and can be set to zero for convenience.
In contrast with the case of a finite harmonic drive, when synchronization requirement is $F\geq|\Delta|R_0$ with $R_0=\sqrt{\gamma_1/2\gamma_2}$, the Arnold tongue remarkably does not depend on the damping parameters  $\gamma_1$ and $\gamma_2$.
This suggests that both in the classical and quantum regime, the Arnold tongue is less susceptible to the adverse effects of noise, making strong entrainment a possibility in the quantum regime.

Now we focus on the differences in frequency entrainment of a harmonically- and squeezing-driven vdP in the deep quantum regime.
\begin{figure}[t]
	\includegraphics[width=\columnwidth]{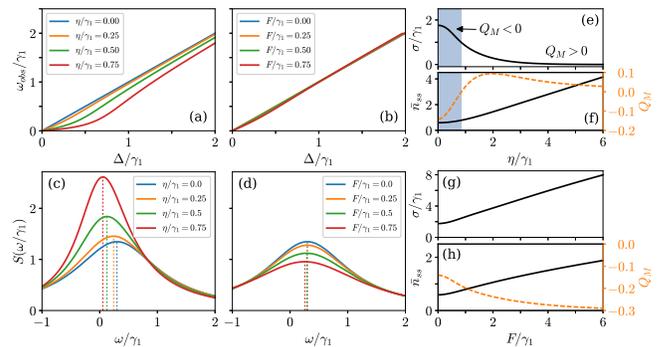}
	\caption{\label{fig:figure3} \textit{Entrainment of squeezing- and harmonically-driven vdP.} Ratio of dissipative processes for all subplots is $\gamma_2/\gamma_1=3$. (a) Four slices of the Arnold tongue at various squeezing parameters $\eta$ show that squeezing produces stronger entrainment when compared with a harmonic drive, shown in (b). (c) and (d) show the power spectrum $S(\omega)$ when $\Delta/\gamma_1=0.3$. Stronger squeezing produces narrower frequency distribution while harmonic drive has the opposite effect and causes broadening. This is highlighted by the solid black lines representing FWHM $\sigma$ of $S(\omega)$ in (e) and (g). The shaded regions of (e) and (f) mark where $Q_M$ (dashed orange line) is negative. (f) Harmonic drive on the other hand produces steady state of vdP for which $Q_M$ is negative for all considered values of $F$.}
\end{figure}
To study frequency entrainment of the oscillator we employ the observed frequency $\omega_{obs}$ defined as the frequency for which power spectrum $S(\omega)$ attains its maximum. When the oscillator is only weakly entrained, $\omega_{obs}$ remains close to the initial detuning $\Delta$. For strong entrainment $\omega_{obs}$ shifts towards $\omega=0$ as the system now oscillates at a frequency close to the external drive.

In Fig.~\ref{fig:figure3}(a)-(b) we compare the observed frequency $\omega_{obs}$ for a squeezing- and harmonically-driven vdP, respectively.
We observe that squeezing produces stronger entrainment, which can be explained by the fact that the Arnold tongue is independent of the damping rates. In the case of a harmonic drive, we observe virtually no frequency entrainment even for very small values of detuning $\Delta$ as already noted in \cite{walter2014quantum}.
This is because the harmonic drive is too weak to overcome the noise inherently present in the vdP and counteracting the drive's efforts to entrain it.
Behavior of the power spectrum $S(\omega)$ is displayed in Fig.~\ref{fig:figure3}(c)-(d).
Stronger squeezing $\eta$ produces sharper distribution of frequencies leading to `cleaner' frequency entrainment as quantified by the spectrum's vanishing full width at half maximum $\sigma$, plotted in Fig.~\ref{fig:figure3}(e).
This is in contrast to the harmonic drive, which has the opposite effect. Here stronger driving $F$ produces frequency distributions with increasing $\sigma$, as shown in Fig.~\ref{fig:figure3}(g).
Finally, we note that both types of driving are capable of producing nonclassical steady states of the oscillator mode as witnessed by the Mandel $Q_M$ parameter, defined in the steady state as $Q_M=[(\Delta\hat{n})^2_{ss} - \bar{n}]/\bar{n}$ \cite{mandel1979sub}, where $\hat{n}=\hat{b}^{\dagger}\hat{b}$ and $\bar{n}=\langle\hat{n}\rangle_{ss}$.
Negativity of $Q_M$ is a sufficient condition for the field to have sub-Poissonian photon number statistics while for $Q_M>0$ no conclusion about nonclassicality can be drawn.

\emph{Experimental realizations.---}
We outline two experimental implementations using trapped ions and an optomechanical setup.
Two implementations using ion traps have been proposed in \cite{lee2013quantum,hush2015spin}.
We follow the approach of \cite{hush2015spin} where the oscillator mode $\hat{b}$ represents a linearly damped motional degree of freedom of the trapped ion.
This linear damping is implemented using standard laser cooling techniques \cite{cirac1992laser}.
The internal degree of freedom of the ion is driven by a standing-wave laser field with Rabi frequency resonant with the first blue sideband transition.
In the Lamb-Dicke regime and when the trapping potentials are tight, this implements an undriven vdP as witnessed by the characteristic ring-shaped steady-state Wigner function pictured in \cite{hush2015spin} and in Fig.~\ref{fig:figure2}(a).
Squeezing can be implemented by an array of techniques such as a combination of standing- and travelling-laser fields \cite{cirac1993dark}, by adiabatically dropping the trap's spring constant \cite{heinzen1990quantum} or by irradiating the ion by two Raman beams separated in frequency by $2\omega_d$ \cite{meekhof1996generation}.

The van der Pol oscillator can also be implemented in a system containing second-order nonlinearity such as the ``membrane-in-the-middle" system \cite{jayich2008dispersive}. We consider a high quality factor membrane where the mechanical dissipation is small. The linear pumping Lindbladian is equivalent to applying a blue-detuned laser by one mechanical frequency whereas the nonlinear damping can be created by applying a laser red-detuned by two mechanical frequencies. The driving force in Eq.~(\ref{sq_H}) can be applied using an electric field gradient created near the membrane. The squeezing can also be generated electrically by modulating the spring constant at twice the mechanical frequency \cite{rugar1991mechanical}.

\emph{Discussion.---} 
In this manuscript, we considered the important problem of noise in the quantum regime of the vdP oscillator.
We demonstrated the control of the dynamics by introducing a squeezing Hamiltonian that counteracts the adverse effects of the noise while maintaining the interesting features of synchronization.
This follows important work showing that including quantum effects can either have favorable \cite{lorch2016genuine} or deleterious effects \cite{lorch2017quantum} on the quality of synchronization.
Our analysis shows that the coherent drive can be replaced by a squeezing drive producing stronger entrainment and a better quality of synchronization as measured by the FWHM of the power spectra.
Finally, following the original proposals by \cite{hush2015spin,walter2014quantum}, we also propose an ion trap and optomechanical implementations of the squeezing-driven vdP. 

Generalizing this idea, consider a network of $N$ self-sustaining oscillators which are weakly coupled and in the synchronization regime. Once synchronized, there exists a critical maximum number of oscillators that can be unsynchronized  such that after transient behavior, they may resynchronize with the network \cite{mirollo1990amplitude}. Owing to stronger frequency entrainment in the quantum regime, the existence of a small number of squeezing driven oscillators in the network could help stabilize the whole network against such disruptions better than the harmonically driven analogue. Such squeezing driven vdPs could herald quantum technologies inspired by synchronization in the quantum regime.

\acknowledgements
Centre for Quantum Technologies is a Research Centre of Excellence funded by the Ministry of Education and the National Research Foundation of Singapore.
This research is supported by the National Research Foundation, Prime Minister's Office, Singapore under its Competitive Research Programme (CRP Award No. NRF-CRP14-2014-02).
SV acknowledges support from an IITB-IRCC grant number 16IRCCSG019.
MH acknowledges insightful discussions with Andy Chia.

%

\end{document}